\documentclass[12pt]{article}
\usepackage[english]{babel}
\usepackage{graphicx,epsfig}
\makeindex \textwidth 170mm \textheight 220mm \topmargin -5mm
\newcommand{\be}{\begin{equation}}
\newcommand{\ee}{\end{equation}}
\newcommand{\bea}{\begin{eqnarray}}
\newcommand{\eea}{\end{eqnarray}}

\def\bm#1{{\mbox{\boldmath$#1$\unboldmath}}}
\def\rfr#1{eq. (\ref{#1})}

\def\bb{\bibitem}
\def\eqi{\begin{equation}}
\def\eqf{\end{equation}}
\def\eqia{\begin{eqnarray}}
\def\eqfa{\end{eqnarray}}
\def\btab{\begin{tabular}}
\def\etab{\end{tabular}}
\def\bar{\begin{array}}
\def\ear{\end{array}}

\def\leti{Lense--Thirring}

\def\lb#1{\label{#1}}

\def\rp#1#2{{#1\over#2}}

\begin{document}
\begin{titlepage}
\begin{flushright}
\today\\
BARI-TH/00\\
\end{flushright}
\vspace{.5cm}
\begin{center}
{\LARGE On the possibility of measuring the Earth's
gravitomagnetic force in a new laboratory experiment }
\vspace{1.0cm}
\quad\\
{Lorenzo Iorio$^{\dag}$\\ \vspace{0.1cm}
\quad\\
{\dag}Dipartimento di Fisica dell' Universit{\`{a}} di Bari, via
Amendola 173, 70126, Bari, Italy\\
\vspace{0.2cm}} \vspace{1.0cm}

{\bf Abstract\\}
\end{center}

{\noindent \small  In this paper we propose, in a preliminary way,
a new Earth--based laboratory experiment aimed to the detection of
the gravitomagnetic field of the Earth. It consists of the
measurement of the difference of the circular frequencies of two
rotators moving along identical circular paths, but in opposite
directions, on a horizontal friction--free plane in a vacuum
chamber placed at South Pole. The accuracy of our knowledge of the
Earth's rotation from VLBI and the possibility of measuring the
rotators'periods over many revolutions should allow for the
feasibility of the proposed experiment.}
\end{titlepage} \newpage \pagestyle{myheadings} \setcounter{page}{1}
\vspace{0.2cm} \baselineskip 14pt

\setcounter{footnote}{0}
\setlength{\baselineskip}{1.5\baselineskip}
\renewcommand{\theequation}{\mbox{$\arabic{equation}$}}
\noindent

\section{Introduction}
In the weak-field and slow-motion approximation of General
Relativity, a test particle in the gravitational field of a slowly
rotating body of mass $M$ and angular momentum ${\bm J}$, assumed
to be constant, is acted upon by a non-central acceleration of the
form (Ciufolini and Wheeler, 1995; Ruggiero and Tartaglia,
2002)\eqi{\bm a}_{\rm GM}=\frac{{\bm v}}{c}\times {\bm B}_{\rm
g}\lb{eqb},\eqf in which ${\bm v}$ is the velocity of the test
particle, $c$ is the speed of light in vacuum and ${\bm B}_{\rm
g}$ is the gravitomagnetic field given by \eqi{\bm B}_{\rm
g}=\frac{2G}{c}\frac{\left[{\bm J}-3\left({\bm J}\cdot
\bm{\hat{r}}\right)\bm{\hat{r}}\right]}{r^3}\lb{gmf}.\eqf In it
$\bm{\hat{r}}$ is the unit position vector of the test particle
and $G$ is the Newtonian gravitational constant.

For a freely orbiting test particle \rfr{eqb} induces on its orbit
the so called Lense--Thirring drag of inertial frames (Lense and
Thirring, 1918; Ciufolini and Wheeler, 1995). Such general
relativistic spin--orbit effect has been experimentally checked
for the first time by analyzing the laser--ranged data to the
LAGEOS and LAGEOS II artificial satellites in the gravitational
field of the Earth with a claimed accuracy of the order of $20\%$
(Ciufolini et al., 1998). The use of the proposed LARES satellite
would greatly increase the accuracy of such space--based
measurement (Ciufolini, 1986; 1998; Iorio et al., 2002a). The
famous GP-B mission (Everitt, 2001), whose flight is scheduled for
the beginning of 2003, is aimed to the detection, among other
things, of a general relativistic spin--spin effect on the
orientation of four spaceborne gyroscopes. The claimed accuracy is
of the order of 1$\%$.

In regard to the possibility of measuring the terrestrial
gravitomagnetic field in an Earth--based laboratory experiment,
many experiments have been proposed\footnote{For a review see
chapter 6 of (Ciufolini and Wheeler, 1995) and (Ruggiero and
Tartaglia, 2002) and references therein. See also (Braginsky et
al., 1977).}, but, up to now, none of them has been practically
implemented due to relevant practical difficulties. Maybe the most
famous of them involves the detection of the gravitomagnetic
precession of the swinging plane of a Foucault pendulum at the
South Pole (Braginsky et al., 1984). Such proposal would not be
easy to practically be implemented due to many sources of errors
recently re--analyzed in (Pascual-S$\acute{\rm a}$nchez, 2002).
Also the proposal by (Cerdonio et al., 1988) should be mentioned.
It is based on an off-line comparison between an astrometric
measurement of the Earth's angular velocity and an inertial
measurement of the angular velocity of the laboratory. Recently,
in (Tartaglia and Ruggiero, 2002) a proposal for detecting the
terrestrial gravitomagnetic field by means of electromagnetic
waves in a Michelson--Moreley type experiment has been put forth.
The effect, in terms of interferometric fringe shift, is really
quite small; however, the advances in technology related to the
gravitational wave detectors like LIGO and VIRGO might allow for a
detection of such effect in future.

Very recently, the influence of a phenomenological gravitational
spin--spin coupling on the free fall of mechanical gyroscopes in
the terrestrial gravitational field which might violate the
equivalence principle (Zhang et al., 2001) has been experimentally
investigated (Luo et al., 2002; Zhou et al., 2002) in some
laboratory preliminary tests. According to them, it seems that
there is no violation of the equivalence principle for extended
rotating bodies at the level of $10^{-7}$.

In this paper we intend to present a possible new Earth--based
laboratory experiment which exploits, in a certain sense, the
concept of the gravitomagnetic clock effect of two
counter--orbiting test particles along identical circular orbits
(Iorio et al., 2002b).
\section{An Earth-based gravitomagnetic clock effect}
According to the gravitational analogue of the Larmor theorem
(Mashhoon, 1993), we could obtain \rfr{eqb} by considering an
accelerated frame rotating with angular velocity \eqi
{\bm\Omega}_{\rm LT}= \frac{{\bm B}_{\rm g}}{2c}.\eqf Indeed, in
it an inertial Coriolis acceleration \eqi{\bm a}_{\rm Cor}=2{\bm
v}\times{\bm\Omega}_{\rm LT}\lb{corio}\eqf is experienced by the
proof mass. It is the same acceleration which, among other things,
induces the Lense--Thirring precession of the swinging plane of
the Foucault pendulum in the experiment proposed by Braginsky and
coworkers (Braginsky et al., 1984) by means of the component of
${\bm\Omega}_{\rm LT}$ along the local vertical direction.

Let us choose an horizontal plane at, say, South Pole\footnote{Of
course, it is easier to prepare an experimental setup in the
Antarctic continent rather than in the Arctic floe. Thanks to the
small size of the proposed apparatus, it should not be too
difficult to find an Antarctic region free enough from seismic
noise and other geological disturbances. Last but not least,
several scientific stations already exist in the Antarctic
continent.}: here the Earth's angular velocity vector
${{\bm\omega}}_{\oplus}$ and ${{\bm\Omega}}_{\rm LT}$ are
perpendicular to it and have opposite directions. Let us choose as
unit vector for the $z$ axis the unit vector
$\bm{\hat{\Omega}}_{\rm LT}$, so that
\bea {{\bm\Omega}}_{\rm LT} & = & \rp{2GJ}{c^2 R_{\rm p}^3}\bm{\hat{z}},\\
{{\bm\omega}}_{\oplus} & = & -\omega_{\oplus}\bm{\hat{z}},\\
{\bm g} & = & -g\bm{\hat{z}},\eea where  ${\bm J}$ is the proper
angular momentum of the Earth\footnote{In fact, while ${\bm J}$
remains constant, the Earth's angular velocity vector
${{\bm\omega}}_{\oplus}$ does not (Bertotti and Farinella, 1990).
Indeed, among other things, it moves around ${\bm J}$ with an
approximate period of 14 months (the Chandler wobble) due to the
oblateness of the Earth (motion with respect to the terrestrial
reference frame). Moreover, there are also the secular precession
of the equinoxes induced by the external lunisolar torque on the
equatorial bulge with a period of 26,000 years and other faster
variations (motion with respect to the celestial reference frame).
The angular velocity of the Earth $\omega_{\oplus}$ suffers a
secular deceleration due to the lunar torque so that the length of
day increases in a year by about $2\times 10^{-5}$ s. In addition,
we have other changes in $\omega_{\oplus}$ over shorter time
scales due to moving masses within and on the Earth, in particular
the oceans and the atmosphere. However, over the characteristic
time scales of the experiment all such variations of
${{\bm\omega}}_{\oplus}$ can be neglected.}, $R_{\rm p}$ is the
Earth polar radius and {\bm g} is the gravitoelectric acceleration
to which, at the poles, the centrifugal acceleration does not
contribute.

A particle which moves with velocity {\bm v} in such polar
horizontal plane is acted upon by the Coriolis inertial force
induced by the noninertiality of the terrestrial reference
frame\footnote{Of course, the centrifugal force is absent at the
Earth's Poles.} and also by the gravitational force of
\rfr{corio}. Such forces have the same line of action and opposite
directions: in an horizontal plane at South Pole the resultant
acceleration is $2v\tilde{\Omega}\equiv 2v(\Omega_{\rm LT}
-\omega_{\oplus})$ and it lies in the plane orthogonally to ${\bm
v}$.

Let us consider an experimental apparatus consisting of a
friction--free horizontal plane placed in a vacuum chamber. Upon
such a desk a small tungsten mass $m$, tied to a sapphire fiber of
length $l$, tension ${\bm T}$ and fixed at the other extremity, is
put in a circular uniform motion. Indeed, the forces which act on
$m$ are the tension of the wire, the Coriolis inertial force and
the Lense--Thirring gravitational force which are all directed
radially; the weight force ${\bm W}=m{\bm g}$ is balanced by the
the normal reaction ${\bm N}$ of the plane and there are neither
the atmospheric drag nor the friction of the plane.
Let us assume the counterclockwise rotation as positive direction
of motion for $m$. At the equilibrium the equation of motion is
\eqi m\omega_+^2 l=T-2m\omega_+ l\tilde{\Omega},\lb{rio}\eqf where
$\omega_+$ is the angular velocity of the mass $m$ when it rotates
counterclockwise and $l$ is the radius of the circle described by
$m$. If the Earth did not rotate the angular velocity of the
particle would be \eqi\omega_0=\sqrt{\rp{T}{ml}}.\eqf For, say,
$m=100$ g, $T=mg=9.798\times 10^{4}$ dyne and $l=100$ cm
$\omega_0=3.13$ rad s$^{-1}$, $P_0$=2 s. The gravitomagnetic and
the Coriolis forces slightly change such circular frequency. Since
$\omega_0>>\tilde{\Omega}$, from \rfr{rio} it follows for both the
counterclockwise and clockwise directions of rotation
\eqi\omega_\pm=\omega_0\mp\tilde{\Omega},\eqf so that we could
adopt as observable \eqi\Delta\omega\equiv\omega_-
-\omega_+=2\tilde{\Omega}\equiv 2(\Omega_{\rm
LT}-\omega_{\oplus}).\lb{dfr}\eqf Of course, the physical
properties of the sapphire fiber and of the tungsten mass should
not change from a set of rotations in a direction to another set
of rotations in the opposite direction, so to allow an exact
cancellation of $\omega_0$ in \rfr{dfr}.

Since on the Earth's surface at the poles $\Omega_{\rm
LT}=3.4\times 10^{-14}$ rad s$^{-1}$, would the experimental
sensitivity of the sketched apparatus allow to measure such so
tiny effect? If we measure the frequency shift $\Delta\omega$ from
the rotational periods of the mass $m$ we have
\eqi\delta\Omega_{\rm LT}=\rp{\delta(\Delta\omega)^{\rm
exp}+\delta\omega_{\oplus}}{2}\eqf with
\eqi\delta(\Delta\omega)^{\rm exp}=\delta\omega_-^{\rm
exp}+\delta\omega_+^{\rm exp}=2\pi\left[\left(\rp{\delta
P_-}{P^2_-}\right)^{\rm exp}+\left(\rp{\delta
P_+}{P^2_+}\right)^{\rm exp}\right].\lb{accu}\eqf

The Earth's angular velocity $\omega_{\oplus}$ is very well known
in a kinematically, dynamically independent way from the Very Long
Baseline Interferometry (VLBI) technique with an accuracy of the
order of\footnote{See on the WEB: {\bf
http://www.iers.org/iers/products/eop/long.html} and {\bf
http://hpiers.obspm.fr/eop-pc/}. See also (Bertotti and Farinella,
1990) in which an accuracy of the order of 2 mas yr$^{-1}\sim
10^{-16}$ rad s$^{-1}$ is arguable.} $\delta\omega_{\oplus }\sim
10^{-18}$ rad s$^{-1}$. In fact, $\omega_{\oplus}$ is not exactly
uniform and experiences rather irregular changes which are
monitored in terms of Length-Of-Day (LOD) by the Bureau
Internationale des Poids et Mesures-Time Section (BIMP) on a
continuous basis\footnote{For these topics see on the WEB: {\bf
http://einstein.gge.unb.ca/tutorial/} and {\bf
http://hpiers.obspm.fr/eop-pc/}. See also (Bertotti and Farinella,
1990).}. Such changes are of the order of
$\Delta\omega_{\oplus}\sim 0.25$ milliarcseconds per year (mas
yr$^{-1})=3.8\times 10^{-17}$ rad s$^{-1}$, so that they are
negligible. A possible source of error might come from our
uncertainty in the position of the proposed polar set--up with
respect to the Earth's crust, i.e. from the polar motion of the
instantaneous axis of rotation of the Earth
$\bm{\hat{\omega}}_{\oplus}$ in terms of the small angles $x$ and
$y$. It turns out that this phenomenon has three components: a
free oscillation with a (measured) period of 435 days (Chandler
Wobble) and an amplitude of less than 1 arcsecond (asec), an
annual oscillation forced by the seasonal displacement of the air
and water masses of the order of 10$^{-1}$ asecs and an irregular
drift of the order of some asecs. There are also some diurnal and
semi--diurnal tidally induced oscillations with an amplitude less
than 1 mas. As a consequence, the position of the pole is unknown
at a level of some meters. The small size of the apparatus should
overcome such problem. Moreover, it can be easily seen that the
impact of such offsets of $\bm{\hat{\omega}}_{\oplus}$ on the
Coriolis force is $2v\omega_{\oplus}\cos\delta$ with $\delta$ of
the order of some asecs or less, so that it is negligible.

In regard to the experimental measurement of the periods
$P_{\pm}$, it should be possible to strongly constrain \rfr{accu}
by choosing suitably the parameters of the apparatus so to
increase the periods and/or by measuring them after many
revolutions. 
\section{Discussion and Conclusions}
In this paper we have proposed a new experiment for measuring the
gravitomagnetic Lense--Thirring effect in an Earth--based
laboratory set--up.

The key point consists of the measurement of the difference of the
rotational frequencies of two test bodies which rotates uniformly
along the same circular paths, but in opposite directions, in an
horizontal, friction--free plane in a vacuum chamber at the South
Pole. The value of the Earth's daily rotation rate has to be
subtracted from such quantity, but the great accuracy in our
knowledge of it from VLBI would allow to single out the tiny
relativistic effect. Over many revolutions it should be possible
to experimentally measure the difference of the rotational
frequencies from the periods to a sufficiently high level of
accuracy to allow for an extraction of the investigated
gravitomagnetic effect.

Of course, many practical difficulties would make the proposed
measurement very hard to be implemented. For example, it turns out
that the the friction force of the plane should be less than
2$\times 10^{-9}$ dyne. Moreover, in order to reach the quoted
accuracy in measuring $\omega_{\oplus}$ with VLBI several years of
continuous observation would be required.

\end{document}